\documentclass[a4paper]{jpconf}
\usepackage{graphicx}
\begin{document}
\title{Connection Between Optical and VHE Gamma-ray Emission in Blazar Jets}

\author{Riho Reinthal*, Elina J. Lindfors*, Daniel Mazin$^\dag$, Kari Nilsson*$^{,\diamondsuit}$, Leo O. Takalo*, Aimo Sillanp\"a\"a* and Andrei Berdyugin*$^{,\diamondsuit}$ on behalf of the MAGIC collaboration \newline\newline
\textit{*Tuorla Observatory, Department of Physics and Astronomy, University of Turku, 21500 Piikki\"o, Finland\\
$^\dag$Institut de Fisica d'Altes Energies (IFAE), E-08193 Bellaterra (Barcelona), Spain\\
$^\diamondsuit$Finnish Centre for Astronomy (FINCA), University of Turku, Finland}}

\address{V\"ais\"al\"antie 20, FI-21500 Piikki\"o, Finland}

\ead{rirein@utu.fi}

\begin{abstract}
MAGIC has been performing optically triggered Target of Opportunity observations of flaring blazars since the beginning of its scientific operations. The alerts of flaring blazars originate from Tuorla Blazar Monitoring Programme, which started the optical monitoring of candidtate TeV blazars in 2002 and has now collected up to eight years of data on more than 60 blazars. These ToO observations have resulted in the discovery of five new VHE $\gamma$-ray emitting blazars (S5 0716+714, 1ES 1011+496, Mrk 180, ON 325 and B3 2247+381). In addition part of the discovery of BL Lac and the discovery of 3C 279 was made during a high optical state. In this contribution we present a detailed analysis of the optical light curves which are then compared to MAGIC observations of the same sources. We aim to answer the question: "Is there a connection between optical and VHE $\gamma$-ray high states in blazars or have we just been lucky?"

\end{abstract}

\section{Introduction}

Blazars are the most extreme case of Active Galactic Nuclei (AGN) in which the relativistic jet originating from close to the central black hole is oriented close to our line of sight. In these objects the dominant radiation component comes from the jet and is non-thermal in origin. Their Spectral Energy Distribution (SED) is characterised by two broad peaks of which the lower energy one is believed to originate from synchrotron emission of the charged particles in the jet. The higher energy one is most commonly explained by inverse Compton scattering of either the synchrotron (Synchrotron Self Compton -- SSC, see e.g. \cite{maraschi92}) or external (External Compton -- EC, \cite{dermer93}, \cite{ghisellini05}) seed photons by the electrons and positrons in the jet, but hadronic models (see e.g. \cite{mannheim93}, \cite{mucke03}) are also considered. Blazars are divided into flat spectrum radio quasars and BL Lacertae objects. The latter are further subdivided into low-, intermediate- and high-frequency peaked BL Lac (LBL, IBL and LBL, respectively) objects according to the frequency of the first peak which typically ranges from near-infrared in LBLs to UV or soft X-rays in HBLs.

Blazars show a variable flux in all wave bands from radio to Very High Energy (VHE, here defined as E$>$100 GeV) $\gamma$-rays on time scales ranging from months to a few minutes. The relationship between the variability at different wave bands appears to be rather complicated. In the simplest case both SSC and EC mechanisms predict a connection between optical and $\gamma$-ray flares. However, there could be time lags between the flares in the different wave bands due to  several mechanisms (see e.g. \cite{sokolov04}, \cite{chatterjee08} for further discussion) with the optical leading the one in $\gamma$-rays or vice versa. Also, in case the synchrotron peak is located far from the optical like in the case of HBLs the variability in the optical band could be very small and the synchrotron flares should be best visible in other wavelengths, usually X-rays. This seems to be the case for the best studied TeV blazars Mrk 421 \cite{lichti08} and Mrk 501 \cite{pian98}. In principle, hadronic models can also explain correlated variability and cannot be excluded as the emission mechanisms of $\gamma$-rays.

MAGIC is performing ToO observations of sources in a high flux state in other wave bands (e.g. optical and/or X-rays and/or high energy $\gamma$-rays). Alerts of optical high states are sent by the Tuorla Blazar Monitoring Program\footnote{http://users.utu.fi/kani/1m} \cite{takalo08}. Observations triggered by optical high states have so far led to the discovery of 5 new sources at VHE $\gamma$-rays. These are, in order of discovery: Mrk 180 \cite{albert06}, 1ES 1011+496 \cite{albert07}, S5 0716+714 \cite{anderhub09}, B3 2247+381 \cite{mariotti10} and ON 325 \cite{mariotti11a}. In addition the discovery of 3C 279 was made and part of the data in the discovery of BL Lac were collected during an optical high state. Also, several $\gamma$-ray flares of known VHE emitters (e.g. the 2007 flare of 3C 279 \cite{aleksic11} and the 2011 flare of 1ES 0806+524 \cite{mariotti11b}) have been detected during optically triggered observations. In this contribution we present an analysis of the optical light curves and compare them to simultaneous MAGIC observations. We discuss the implications of the results to the optical to VHE $\gamma$-ray connection.

\section{Analysis of the optical light curves}

Photometric observations are performed on the 35 cm Celestron telescope attached to the body of the 60 cm KVA telescope, which is located on the Canary island of La Palma, Spain and operated remotely from Tuorla Observatory, and the Tuorla 1 m telescope. Observations are done in the R-band and the magnitudes are measured using differential photometry with calibration stars in the same CCD frame as the object (see \cite{nilsson07} for a more detailed overview). The light curves are available on the project web page$^1$ and are updated on a daily basis.

The monitoring started in 2002 and the original list consisted of 24 candidate TeV blazars chosen from \cite{costamante02} (with $\delta >$ 20$^\circ$ for them to be observable from Tuorla). Over time the number of monitored sources has grown to over 60 including many candidates detected at high energy (HE) $\gamma$-rays by the Fermi telescope. The goal of the monitoring is to study the optical variability of these sources and to determine their baseline flux. We have also studied the host galaxy's contribution to most objects' overall brightness \cite{nilsson07}. This way we can get a direct measurement of the AGN's core brightness by subtracting the host galaxy contribution. For this study we chose 31 of the best sampled light curves including all 24 from the original sample.

\begin{figure}[b]
\includegraphics[width=0.5\linewidth]{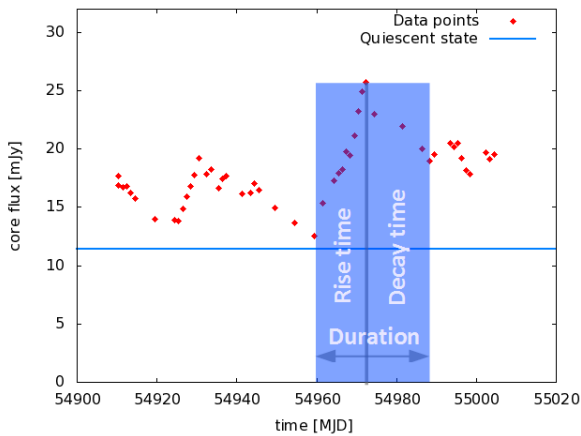}\hspace{2pc}
\begin{minipage}[b]{0.45\linewidth}\caption{\label{zoomedflare}\small{An example of the parts of the flare: peak of the flare is the point in the flare with the highest flux; beginning and end of the flare are the points with the lowest flux prior or after exceeding the flare condition respectively; rise time is the time from the beginning of the flare to the peak and decay time the time from the peak to the end.}}
\end{minipage}
\end{figure}

Historically a source was said to be flaring in the optical if its core flux increased by $\geq$ 50\% compared to its quiescent level. For determining the quiescent level a general rule that a source would spend $\sim$20\% of the time at or below this level was followed. Both the quiescent states and the flares were identified by visual inspection of the light curves. One of the purposes of this study was to calculate exact values for the quiescent states according to some common principles and to automate the identification of flares. As quiescent state we used the weighted average calculated from all the data points except for those belonging to 'flares'. For the purposes of these calculations 'flares' were all data points that were typically more than three (the exact value varied slightly from source to source) standard deviations above the average. As an initial value for the calculations we used the historical value of quiescent state i.e. the flux of which 20\% of the whole data set was of lower level. In most cases the quiescent level ended up around the value where we would have put it by eye and also usually close to the 20\% value used previously.
The relative increase in core flux required for a brightening to be constituted as a flare was derived for each source independently depending on the amplitude of their variability -- the larger the amplitude the higher increase in flux was required:
\begin{math}
FlareCond \sim Qstate * \frac{F_{max}}{F_{min}}
\end{math}
where $FlareCond$ is the flare condition, $Qstate$ the quiescent state and $F_{max}$ and $F_{min}$ are the highest and lowest measured core flux respectively. A total of 132 flares were counted since the beginning of the monitoring of which 108 occurred after MAGIC started regular observations. These numbers are slightly higher than the number of flares identified by eye from the light curves, because: a) if there is a long flare but with a gap in the data, software identifies it as two distinct flares and b) some sources (e.g. OJ 287) show "flaring states" where the flux can drop below the flaring condition briefly but immediately increase again which the software counts as separate flares. We also calculated the rise and decay times of the flares and flare durations where possible (refer to Fig. \ref{zoomedflare} for explanation of these terms). A summary of all the flares of all the sources can be seen in Table \ref{flarestats}.

\begin{table}[h]
\begin{center}
\includegraphics[width=1.\linewidth]{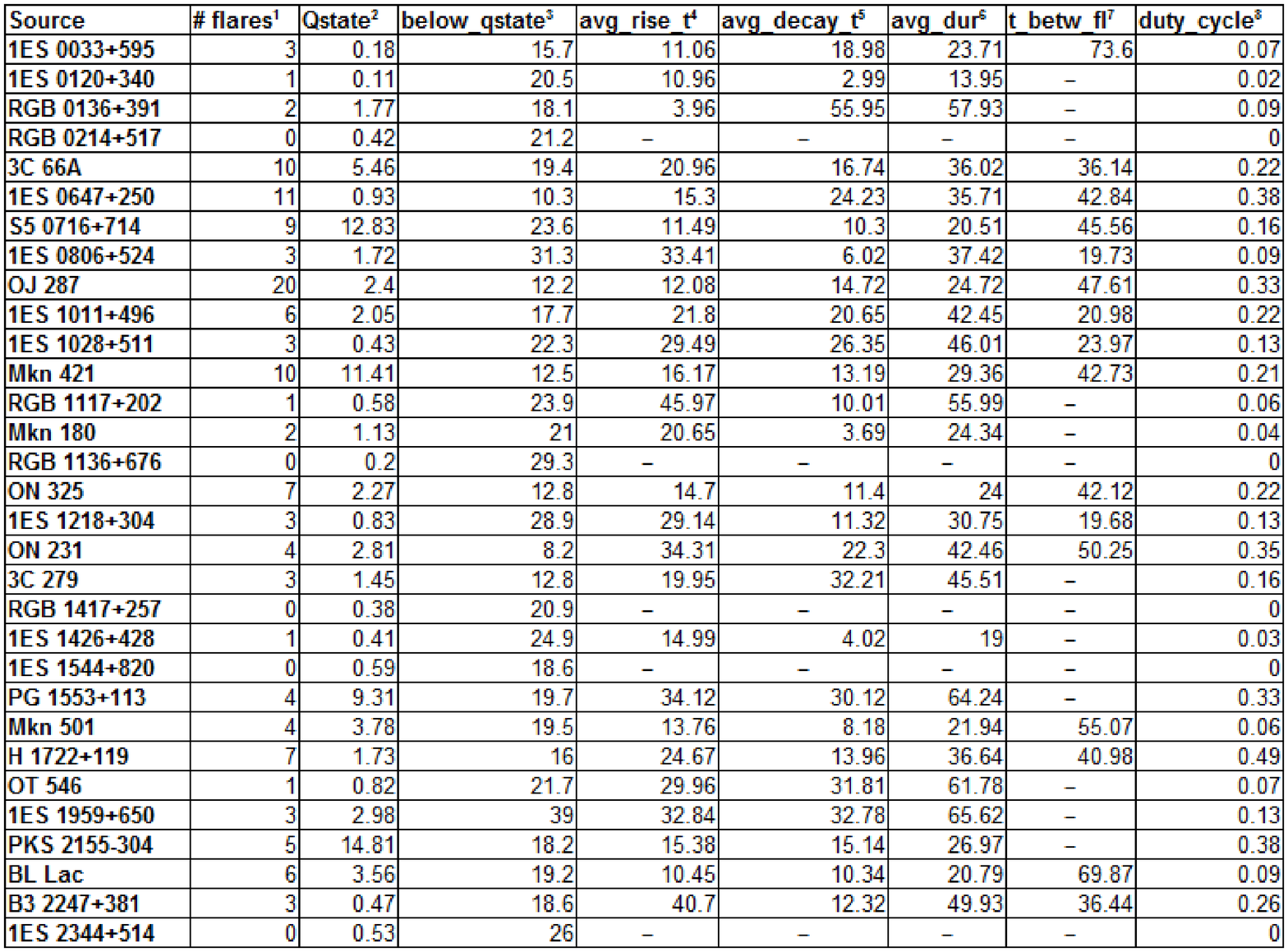}
\end{center}
\caption{\label{flarestats}\small{Results of the light curve analysis source-by-source.\\
$^1$Absolute value of flares identified since the beginning of the programme.\\
$^2$Quiescent state flux in mJy.\\
$^3$Percentage of data points below quiescent state.\\
$^4$Average rise time in days.\\
$^5$Average decay time in days.\\
$^6$Average flare duration in days.\\
$^7$Average time between two flares in the same observation period (the $\sim$6 months when the source is visible).\\
$^8$Estimate of the duty cycle (fraction of time the source spends above flaring condition).\\
Note that for the calculations of the averages only values $>$0 were used. For the calculation of the duty cycles all data gaps longer than three weeks were excluded.}}
\end{table}

\begin{table}
\begin{center}
\includegraphics[width=1.\linewidth]{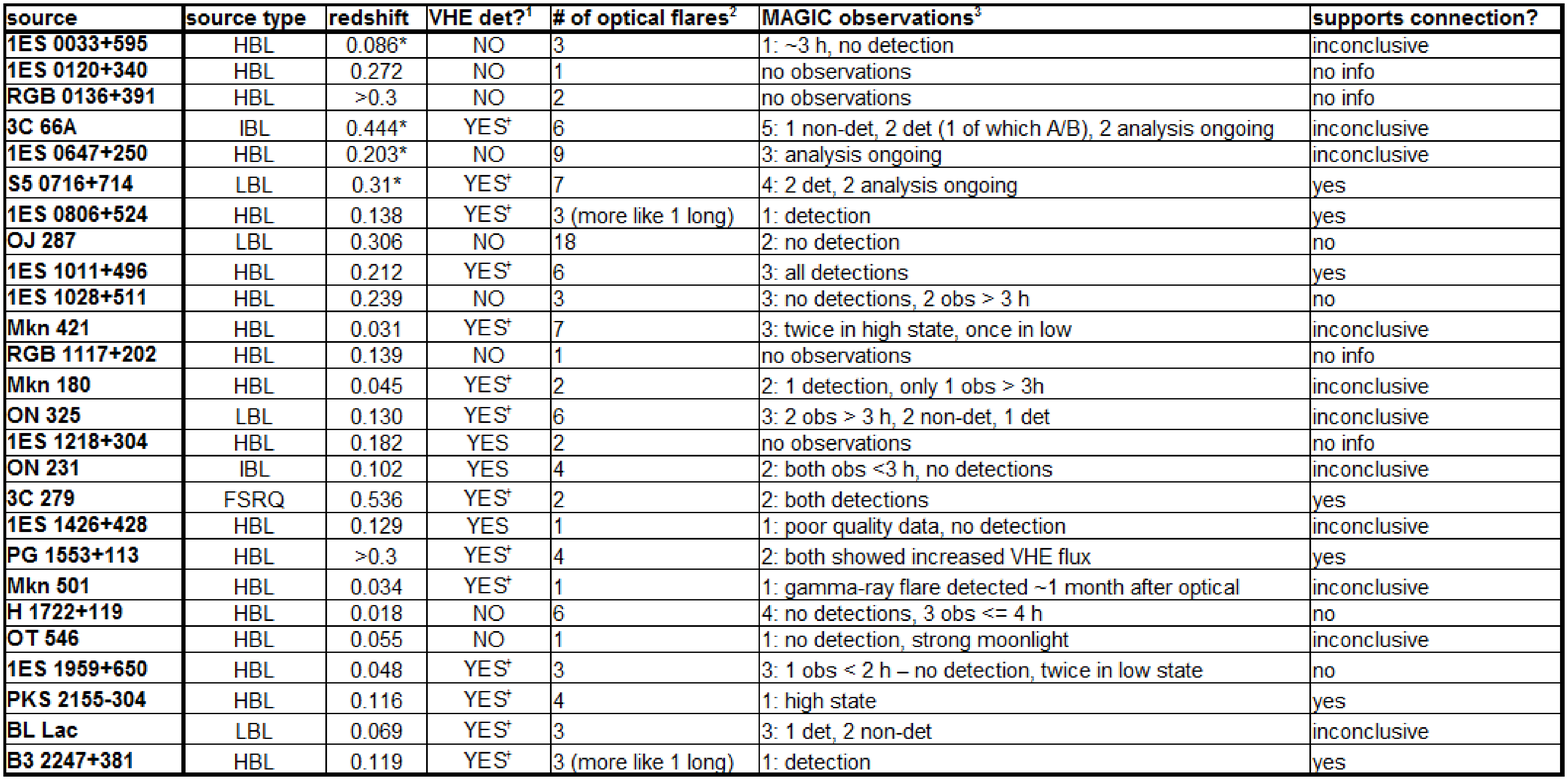}
\end{center}
\caption{\label{summary}\small{Summary of optical flares detected since MAGIC started observations and their comparison to contemporaneous VHE data. The redshifts marked with an asterisk are tentative values.\\
$^1$Flag whether the source has been detected at VHE $\gamma$-rays by MAGIC or any other IACT. The ones seen by MAGIC are marked with a cross.\\
$^2$Number of optical flares since the beginning of Cycle 1 in MAGIC.\\
$^3$The number and short summary of the contemporaneous MAGIC observations.}}
\end{table}

\section{Optical to VHE $\gamma$-ray comparison}

MAGIC (Major Atmospheric Gamma-ray Imaging Cherenkov) is a system of two imaging atmospheric Cherenkov telescopes (IACTs) located on the Canary island of La Palma, Spain at $\sim$2200 m asl. The first telescope started observations in 2004 and since November 2009 the two telescopes have been observing in stereoscopic mode. Both telescopes have 17 m mirror dishes -- the largest of any IACTs -- allowing them to reach lower energy threshold than other instruments of their type. In the stereoscopic mode MAGIC reaches an energy threshold of $\sim$50 GeV ($\sim$25 GeV with a special trigger), sensitivity of $\sim$0.8\% of Crab Nebula flux (5 $\sigma$ in 50 h $>$250 GeV), an energy resolution of $\sim$15-20\% and an angular resolution of 0.07 degrees \cite{albert08} making it the best instrument of its kind, particularly for energies below 200 GeV.

Because MAGIC is performing ToO observations of blazars that are in a high state in the optical there is simultaneous or quasi-simultaneous optical and VHE data for quite many flares. Of the 108 optical flares that occurred after the beginning of Cycle 1 in MAGIC (June 2005) for 56 there is some simultaneous MAGIC data but for only 49 of those there is more than 3 hours of good quality data. This is mostly constrained by the weather and moon conditions as well as conflicts in the schedule. On 27 occasions the source was detected at VHE (on 3 occasions in a low state) -- a total of 14 different sources were seen (note that for Mrk 501 the VHE flare occurred $\sim$1 month after the optical), 7 of them for the first time by MAGIC. In 19 cases the observations resulted in upper limit only and for 8 flares information was not available or the analysis is ongoing. The remaining two observations resulted in the detection of VHE $\gamma$-rays from the 3C 66A/B region \cite{aliu09}. The results of the comparison of the optical and VHE data are summarised in Table \ref{summary}.

There are only two sources, Mrk 421 and Mrk 501, that can be reliably detected at low flux states in reasonable time scales and that have good enough VHE coverage that allows us to study the correlation between optical and VHE $\gamma$-ray states. The long-term monitoring of these sources by MAGIC is summarised in \cite{hsu09}. From the Mrk 421 data set they did not find any correlation between the optical and VHE wave bands whereas Donnarumma \etal \cite{donnarumma09} saw a hint of correlation between the two during the 2008 flare. For both Mrk 421 and 501 extensive data sets have been collected (see e.g. \cite{abdo11a}, \cite{abdo11b}) allowing to study the correlation between optical and VHE $\gamma$-rays in unparalleled detail. However, both of these sources are so-called extreme blazars \cite{costamante01} so they are not really representative of the whole blazar or the HBL classes.
The long-term monitoring of 1ES 1959+650 by MAGIC has always detected the source in a low state \cite{hsu09} whereas in the optical the source has shown three flares since the beginning of the monitoring. Although difficult to observe from the Northern hemisphere due to large zenith angles, PKS 2155-304 can still be reliably detected by MAGIC during strong outbursts. One such was reported by H.E.S.S in July 2006 (coinciding with an optical flare) triggering MAGIC observations and leading to a strong detection of the source (see \cite{hadasch07} and \cite{hadasch09}). Also, H.E.S.S. saw a strong correlation between the optical and VHE bands during the multi-wavelength observations of PKS 2155-304 in August-September 2008 during a low VHE state of the source (see \cite{aharonian09a}) but this correlation does not always seem to be there. The results of five years of observations of PG 1553+113 by MAGIC reported in \cite{aleksic11b} suggest a correlation between the optical and VHE wave bands for this source. Only for a handful of the other sources we have MAGIC observations in both high and low flux states (e.g. S5 0716+714, 1ES 0806+524 and 3C 279) and in most cases there seems to be a connection between activities in the two wavebands but it is difficult to prove with the low statistics at VHE.

\section{Discussion and conclusions}

Optically triggered ToO observations have resulted in the discovery of five new VHE $\gamma$-ray emitting blazars by MAGIC so far. Additionally several other discoveries and $\gamma$-ray flares have coincided with optical high states. If the synchrotron emission and the IC emission arise from the same emission region, a connection between optical and VHE $\gamma$-ray states would be expected. However, in case the synchrotron peak is located far from the optical waveband (which can be the case for HBLs) such connection would be difficult if not impossible to observe due to low amplitude of variability in the optical. And in case of different emission regions or mechanisms there might be no direct connection. It seems that for most sources the optical and VHE $\gamma$-ray states are somewhat connected. However there is at least one example (1ES 1959+650) where this does not seem to be the case which indicates that at least in some sources there are different emission mechanisms at work or that the two emissions originate from different regions.

In order to study the problem more in detail correlation studies between the optical and VHE $\gamma$-ray states are required. This would require more complete VHE light curves with reliable detections of the sources also at lower flux states. Unfortunately with the current generation of IACTs these kinds of studies are only feasible for a few sources and would require a significant fraction of the available observation time to be dedicated to monitoring. Otherwise we have to rely on future instruments to shed more light into this matter.\\
\\
\footnotesize{\textit{Acknowledgements}: The work of RR and EL has been supported by grant 127740 of the Academy of Finland. We would also like to thank the Instituto de Astrof\'{\i}sica de Canarias for the excellent working conditions at the Observatorio del Roque de los Muchachos on La Palma. The support of the German BMBF and MPG, the Italian INFN, the Swiss National Fund SNF, and the Spanish MICINN is gratefully acknowledged. This work was also supported by the Marie Curie program, by the CPAN CSD2007-00042 and MultiDark CSD2009-00064 projects of the Spanish Consolider-Ingenio 2010 programme, by grant DO02-353 of the Bulgarian NSF, by the YIP of the Helmholtz Gemeinschaft, by the DFG Cluster of Excellence "Origin and Structure of the Universe", by the DFG Collaborative Research Centers SFB823/C4 and SFB876/C3, and by the Polish MNiSzW grant 745/N-HESS-MAGIC/2010/0.}

\end{document}